\documentclass[12pt,a4paper]{article}

\usepackage{amsmath,amsthm,amssymb}
\usepackage[dvips]{graphicx}

\makeatletter
\renewcommand{\theequation}{\arabic{section}.\arabic{equation}}
\@addtoreset{equation}{section}
\makeatother
\renewcommand\arraystretch{1.4}
\begin{document}

\vspace*{0.5cm}
\begin{center}
{\Large \bf  Inequivalent Quantizations of the $N = 3$  Calogero}

\vspace{.3cm}

{\Large \bf  model with Scale and Mirror-$S_3$ Symmetry}
\end{center}

\vspace{0.5cm}

\begin{center}
Nobuhiro Yonezawa\footnote{
\,e-mail: yonezawa@post.kek.jp}
and Izumi Tsutsui\  \\

\bigskip
\bigskip
{\em Institute of Particle and Nuclear Studies\\
High Energy Accelerator Research Organization (KEK)\\
Tsukuba 305-0801, Japan}
\end{center}

\vspace{2.5cm}

\begin{abstract}
We study the inequivalent quantizations of the $N = 3$ Calogero model by separation of variables, in which the model decomposes into the angular and the radial parts.  
Our inequivalent quantizations respect the \lq mirror-$S_3$\rq\ invariance (which realizes the symmetry under the cyclic permutations of the particles) and the scale invariance in the limit of vanishing harmonic potential.   
We find a two-parameter family of novel quantizations in the angular part and classify the eigenstates in terms of the irreducible representations of the $S_3$ group.   The scale invariance restricts the quantization in the radial part uniquely, except for the eigenstates coupled to the lowest two angular levels for which two types of boundary conditions are allowed independently from all upper levels.
It is also found that the eigenvalues corresponding to the singlet representations of the $S_3$ are universal (parameter-independent) in the family, whereas those corresponding to the doublets of the $S_3$ are dependent on one of the parameters.    These properties are shown to be a consequence of the spectral preserving $SU(2)$ (or its subrgoup $U(1)$) transformations allowed in the family of inequivalent quantizations. 
\end{abstract}

\bigskip

\newpage

\section{Introduction}

The $N$-body Calogero model, which describes $N$ particles interacting each other by the combined inverse square and 
harmonic potential on a line, is exactly solvable and yet admits a diversity of mathematical extensions and physical applications.    For the mathematical side, Calogero's analysis \cite{Calogero1} for $N = 3$  has been extended to general $N$ \cite{Calogero2, Calogero3} as well as to models with modified potentials \cite{Calogero-Marchioro, Sutherland} and Lie-algebraic structures \cite{Olshanetsky-Perelomov} (see also \cite{Polychronakos1, Brink-Hansson-Konstein-Vasiliev, Sogo, Diejen, Gomez-Ullate-Gonzalez-Lopez-Rodriguez, Diejen-Vinet} and the references therein).  For the physical side, its relevance to high energy physics has been argued, albeit in specific circumstances, in various areas,  {\it e.g.}, in the Yang-Mills theory \cite{Gorsky-Nekrasov}, two-dimensional QCD \cite{Niemi-Pasanen}, superstrings \cite{Ferretti} and black holes \cite{Gibbons-Townsend}.   Moreover, its potential application has also been argued in condensed matter physics, {\it e.g.}, for spin chains \cite{Polychronakos2}, quantum Hall effect \cite{Azuma-Iso}, magnons \cite{Inozemtsev} and the electron-hole interaction \cite{Markvardsen-Johnson}.

In considering the application to particles with anyon statistics, Veigy \cite{Veigy} made an important observation that, when the coupling constant of the interaction lies in a certain range,  the Calogero model admits a wider class of solutions than those obtained by Calogero.    Technically, the possibility of the extension is found in the treatment of the singularities in the inverse square potential, where a specific (the Dirichlet) boundary condition has conventionally  been adopted, allowing for only bosons and fermions for the possible statistics of the particles.  
It is shown in \cite{Veigy} that, if one adopts a different boundary condition at the singularities  in the {\it angular} part of the model, which arises after the separation of variables is implemented for the $N=3$ case, 
one indeed finds extra eigenstates that correspond neither to bosons nor fermions.   More recently, it has also been shown in \cite{Falomir-Pisani-Wipf, Basu-Mallick-Ghosh-Gupta, Birmingham-Gupta-Sen} that the model admits novel solutions if we consider a general class of boundary conditions for the {\it radial} part, even if the conventional boundary condition is adopted for the angular part.  

To explore a fuller class of solutions available in the Calogero model, which amounts to exploring the possible {\it inequivalent quantizations} of the model,   it is necessary to study the combined extension in the boundary conditions both in the angular and radial parts in more general terms, and this has been initiated in \cite{Tsutsui2} for $N=3$.   For the angular part, the class of boundary conditions considered in \cite{Tsutsui2} forms a two-parameter family (containing the cases treated by Calogero and  Veigy earlier)
and respects a dihedral $D_6$ symmetry, which derives from the demand of indistinguishability of the particles.  We found that the model can still be solvable, but the spectrum cannot be obtained explicitly except for a number of special cases such as the ones considered earlier.    The purpose of the present paper is to provide another class of solutions 
allowed when we relax the demand from the $D_6$ symmetry to its subgroup $S_3$ and impose instead scale symmetry in the vanishing harmonic potential limit.  
In physical terms, the $S_3$ symmetry is equivalent to the invariance under cyclic permutations of the particles, while the 
additional scale symmetry ensures the smooth limit to the pure inverse square potential system at the quantum level.   For the angular part, 
this leads to a novel two-parameter family of inequivalent quantizations, where the spectra can be obtained in closed form as well as the explicit solutions classified according to the irreducible representations of the group $S_3$.   For the radial part, on the other hand, the scale symmetry specifies the quantization essentially uniquely to one obtained under the Dirichlet boundary condition, except for the eigenstates coupled to the 
lowest two angular levels for which the Neumann  boundary condition is permitted as well.  This allows us to have a number of possible combinations of boundary conditions for the angular and radial parts.   We also observe that the levels which are singlets of $S_3$ are universal, {\it i.e.}, independent of the parameters of the family, whereas the doublets of $S_3$ are dependent on one of the parameters.  This will be seen as a consequence of the spectral preserving $SU(2)$ (or its subgroup $U(1)$) transformations allowed in the family of inequivalent quantizations of the model. 

The plan of the paper is as follows.   We recapitulate in section 2 the procedure of inequivalent quantizations of the $N=3$ Calogero model by separation of variables and present the symmetries we impose.   We give a detailed discussion on the eigenstates and eigenvalues in the angular part in section 3.  These results are then combined with the counterparts of the radial part discussed in section 4 to provide the solutions and the spectra of the entire system.   Section 5 is devoted to the discussion of the universality of the $S_3$-singlets  based on the spectral preserving $su(2)$ algebra found earlier  \cite{Tsutsui3}.   Finally, our conclusion is presented in section 5.

\section{The $N=3$ Calogero model}

In this section we provide a framework for quantizing the $N=3$ Calogero model based on the separation of variables method, in which we split the model into the radial and angular parts.
Concerning this, there are two important issues that require a particular care, one of which is the treatment of the singularity in the potential and the other is the choice of the symmetry we shall adopt for our inequivalent quantizations.  

\subsection{Separation of variables}

The $N$ body Calogero model \cite{Calogero2, Calogero3} is a system of 
$N$ particles 
governed by 
the Hamiltonian,
\begin{eqnarray}
H = -\frac{\hbar^2}{2m}\sum _{i=1}^N \frac{\partial ^2 }{\partial x _i^2}
+ \sum_{j=1}^N \sum _{i=j+1}^N \Bigl\{ \frac{g}{(x_i-x_j)^2}
+\frac{1}{4} m \omega ^2 (x_i-x_j)^2 \Bigr\},
\label{nham}
\end{eqnarray}
where $x_i$, $i = 1, \ldots, N$ represent the positions of the particles under interaction.
As is well-known, the system can be analyzed by the method of separation of variables  (see,~{\it e.g.},~\cite{Tsutsui2}), which for  
the $N=3$ case  begins with the 
use the Jacobi coordinates,
\begin{align}
x =\frac{x_1-x_2}{\sqrt{2}},\hspace{10mm}
y =\frac{x_1+x_2-2 x_3}{\sqrt{6}},\hspace{10mm}
z =\frac{x_1+x_2+x_3}{\sqrt{3}}\label{Jacobi coordinate},
\end{align}
and passes on to the polar coordinates $(x, y) = (r \sin \phi, r \cos \phi)$.  
In the unit $2m=1$, $\hbar=1$,  this procedure brings the Hamiltonian to the separable form,
\begin{align}
H = -\frac{\partial ^2 }{\partial r ^2}
-\frac{1}{r}\frac{\partial  }{\partial r }
-\frac{\partial ^2 }{\partial z^2}
+\frac{3}{8} \omega ^2 r^2
+\frac{1}{r^2} \Bigl\{  -\frac{\partial ^2 }{\partial \phi ^2}
+ \frac{g}{2} \frac{9}{ \sin ^2 3 \phi} \Bigr \}.
\label{threeham}
\end{align}

It is also known that, if $g<-\frac{1}{2}$, the model does not admit a ground state \cite{L.D.Landau}, and
if $\frac{3}{2} \le g$, the Hamiltonian $H$ is essentially self-adjoint and leads to a unique quantization.
In the present paper, we consider the cases  
$-\frac{1}{2} < g<\frac{3}{2}$, $g \ne 0$,
where $H$ admits self-adjoint extensions allowing for inequivalent quantizations.  Here,  the two particular points $g=0$ and $g=-\frac{1}{2}$ are excluded because the model becomes a harmonic oscillator in the former case while it 
requires an independent treatment technically in 
the latter case.   For our later convenience, 
instead of $g$ we use $\nu$ introduced as
\begin{align}
g=2\nu(\nu-1),  \hspace{10mm}  
\frac{1}{2} < \nu < \frac{3}{2}, \hspace{10mm} 
\nu \ne 1.
\end{align}

On account of the separable form (\ref{threeham}), we have  the Hamiltonian $H = H_0 + H_{\rm rel}$ with $H_0$ describing the centre of mass system and  $H_{\rm rel}$ the relative motion.  The latter splits further 
into $H_{\rm rel} = H_r + H_\Omega$, where $H_r$ and $H_\Omega$ are the Hamiltonians for the radial and angular parts, respectively.  
Consider now the angular eigenvalue equation, 
\begin{align}
H_\Omega\,\psi_\lambda (\phi) = \lambda \,\psi_\lambda (\phi), 
\hspace{10mm} 
H_\Omega  =-\frac{d^2}{d\phi^2}+\frac{9\nu(\nu-1)}{\sin^2 3 \phi},
\label{angular equation}
\end{align}
with eigenvalue $\lambda$.  To each $\lambda$, we further  consider
the effective radial Hamiltonian $H_{r, \lambda} = H_r + \lambda/r^2$ and its eigenvalue equation,
\begin{align}
H_{r,\lambda}  \psi_{E} (r; \lambda)&= E\, \psi_{E} (r;\lambda),
\hspace{10mm} 
H_{r,\lambda} =
 -\frac{d^2}{dr^2}-\frac{1}{r}\frac{d}{dr}+\frac{3}{8}\omega^2 r^2 + \frac{\lambda}{r^2}.
\label{radial equation}
\end{align}
The eigenfunction for the entire system (apart from that for the centre-of-mass system in the $z$-direction) with total energy $E$ is obtained as $\psi_{E} (r; \lambda) \psi_\lambda (\phi)$ from the solutions of these equations (\ref{angular equation}) and (\ref{radial equation}).

\subsection{Singularities and formal symmetries}

In our quantization, an important point to note is that both of  the angular Hamiltonian $H_\Omega$ and the radial Hamiltonian $H_{r,\lambda}$ are ill-defined at  the singular points of the potentials, {\it i.e.}, at 
$\phi$ for which $\sin 3\phi = 0$ and at $r = 0$.  
Thus, to quantize the model properly and solve the eigenvalue equations, we need to consider self-adjoint extensions to each of the operators, that is, we provide an appropriate domain of definition in such a way that the entire Hamiltonian $H$ be self-adjoint.  In practice, this is accomplished
by furnishing a set of connection conditions for the eigenfunctions at these singular points
according to the general scheme for singular Hamiltonians (\ref{radial equation}) and (\ref{angular equation}), and in what follows we adopt the scheme presented in \cite{Tsutsui1}. 

\begin{figure}[t]
  \begin{center}
\unitlength 1pt
\begin{picture}(161.1621,173.4480)(247.8861,-178.5069)
%
\special{pn 8}%
\special{pa 3616 1940}%
\special{pa 5514 828}%
\special{fp}%
%
\special{pn 8}%
\special{pa 4014 416}%
\special{pa 5126 2314}%
\special{dt 0.045}%
%
\special{pn 8}%
\special{pa 4014 416}%
\special{pa 5126 2314}%
\special{dt 0.045}%
%
\special{pn 8}%
\special{pa 4014 416}%
\special{pa 5126 2314}%
\special{dt 0.045}%
%
\special{pn 8}%
\special{pa 4032 2346}%
\special{pa 5108 428}%
\special{dt 0.045}%
\put(392.4261,-57.8160){\makebox(0,0)[lb]{$P_1$}}%
\put(284.3825,-178.1456){\makebox(0,0)[lb]{$R_3$}}%
\put(334.6101,-182.1204){\makebox(0,0)[lb]{$P_3$}}%
\put(284.0211,-25.2945){\makebox(0,0)[lb]{$R_2$}}%
\put(247.8861,-96.8418){\makebox(0,0)[lb]{$R_1$}}%
\put(398.9304,-135.1449){\makebox(0,0)[lb]{$P_2$}}%
%
\special{pn 8}%
\special{pa 3470 1376}%
\special{pa 5660 1376}%
\special{dt 0.045}%
\special{sh 1}%
\special{pa 5660 1376}%
\special{pa 5594 1356}%
\special{pa 5608 1376}%
\special{pa 5594 1396}%
\special{pa 5660 1376}%
\special{fp}%
%
\special{pn 8}%
\special{pa 4576 2470}%
\special{pa 4576 280}%
\special{fp}%
\special{sh 1}%
\special{pa 4576 280}%
\special{pa 4556 348}%
\special{pa 4576 334}%
\special{pa 4596 348}%
\special{pa 4576 280}%
\special{fp}%
%
\special{pn 8}%
\special{ar 4580 1380 678 678  1.5776455 6.2831853}%
\special{ar 4580 1380 678 678  0.0000000 1.5707963}%
%
\special{pn 8}%
\special{pa 3616 1940}%
\special{pa 5514 828}%
\special{fp}%
%
\special{pn 8}%
\special{pa 4014 416}%
\special{pa 5126 2314}%
\special{dt 0.045}%
%
\special{pn 8}%
\special{pa 4014 416}%
\special{pa 5126 2314}%
\special{dt 0.045}%
%
\special{pn 8}%
\special{pa 4014 416}%
\special{pa 5126 2314}%
\special{dt 0.045}%
%
\special{pn 8}%
\special{pa 4032 2346}%
\special{pa 5108 428}%
\special{dt 0.045}%
\put(392.4261,-57.8160){\makebox(0,0)[lb]{$P_1$}}%
\put(284.3825,-178.1456){\makebox(0,0)[lb]{$R_3$}}%
\put(334.6101,-182.1204){\makebox(0,0)[lb]{$P_3$}}%
\put(284.0211,-25.2945){\makebox(0,0)[lb]{$R_2$}}%
\put(247.8861,-96.8418){\makebox(0,0)[lb]{$R_1$}}%
\put(398.9304,-135.1449){\makebox(0,0)[lb]{$P_2$}}%
%
\special{pn 8}%
\special{pa 3470 1376}%
\special{pa 5660 1376}%
\special{dt 0.045}%
\special{sh 1}%
\special{pa 5660 1376}%
\special{pa 5594 1356}%
\special{pa 5608 1376}%
\special{pa 5594 1396}%
\special{pa 5660 1376}%
\special{fp}%
%
\special{pn 8}%
\special{pa 4576 2470}%
\special{pa 4576 280}%
\special{fp}%
\special{sh 1}%
\special{pa 4576 280}%
\special{pa 4556 348}%
\special{pa 4576 334}%
\special{pa 4596 348}%
\special{pa 4576 280}%
\special{fp}%
\put(408.3255,-95.3964){\makebox(0,0)[lb]{$x$}}%
\put(333.8874,-17.3448){\makebox(0,0)[lb]{$y$}}%
%
\special{pn 20}%
\special{ar 4580 2050 20 20  0.0000000 6.2831853}%
%
\special{pn 20}%
\special{ar 5170 1720 20 20  0.0000000 6.2831853}%
%
\special{pn 20}%
\special{ar 3990 1720 20 20  0.0000000 6.2831853}%
%
\special{pn 20}%
\special{ar 4000 1030 20 20  0.0000000 6.2831853}%
%
\special{pn 20}%
\special{ar 4580 700 20 20  0.0000000 6.2831853}%
%
\special{pn 20}%
\special{ar 5160 1030 20 20  0.0000000 6.2831853}%
\put(336.7782,-68.6565){\makebox(0,0)[lb]{$1$}}%
\put(364.9635,-93.9510){\makebox(0,0)[lb]{$2$}}%
\put(354.1230,-130.0860){\makebox(0,0)[lb]{$3$}}%
\put(318.7107,-138.0357){\makebox(0,0)[lb]{$4$}}%
\put(291.2481,-114.1866){\makebox(0,0)[lb]{$5$}}%
\put(302.0886,-78.0516){\makebox(0,0)[lb]{$6$}}%
%
\special{pn 8}%
\special{pa 3620 820}%
\special{pa 5520 1932}%
\special{fp}%
\end{picture}%
  \end{center}
  \caption{Axes of reflections $P_i$, $R_i$, $i = 1, 2, 3$, on the plane $(x, y)$.  The numbers $k = 1, \ldots, 6$ indicate the sectors $s_k = (\phi_{k-1}, \phi_{k})$ defined along the circle.}
  \label{xy-plane}
\end{figure}
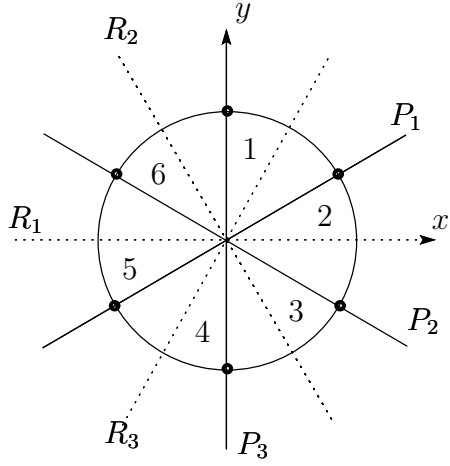

We start with the angular Hamiltonian, and for this we first note the formal symmetry of the operator $H_\Omega $ in  (\ref{angular equation}).   From of the potential, it is clear that $H_\Omega $ is invariant under the reflections,
\begin{align}
P_3:\phi \mapsto -\phi,
\hspace{10mm} 
R_3:\phi \mapsto \frac{\pi}{3}-\phi,
\label{defprity}
\end{align}
and also under other four reflections $P_i$, $R_i$, $i = 1, 2$, analogously defined (see Fig.\ref{xy-plane}). 
It follows that the Hamiltonian is invariant under the rotation by angle ${{\pi}\over 3}$,
\begin{align}
\mathcal{R}_\frac{\pi}{3}=R_3 \circ P_3:\phi \mapsto \phi+\frac{\pi}{3}, 
\end{align}
and hence under the successive ones $(\mathcal{R}_\frac{\pi}{3})^j$,  $j=0,1,2,3,4,5$, as well.
Combining all these reflections and rotations, we obtain
a $D_6$ group as the group of the formal symmetry of $H_\Omega$.    
On the angular wave function $\psi(\phi)$, the action of the operation $g \in D_6$ is realized by 
\begin{align}
\hat g\, \psi (\phi )= \psi (g^{-1} \phi).
\label{defgection}
\end{align}
It should be stressed, however, that the symmetry $D_6$ is still formal and the actual symmetry at the quantum level should be determined by taking account of the connection condition at the singularities.

In order to provide the connection condition at the singular points $\phi= \phi_k$,
\begin{align}
 \phi_k = \frac{k\pi}{3},
 \hspace{10mm} 
 k= 0, 1,2,3,4,5,
\label{singpoints}
\end{align}
we first focus on the singularity at $\phi = \phi_0 = 0$ and introduce the boundary vectors 
\begin{align}
B_{0}(\psi) :=
\left(
  \begin{array}{c}
     W \left[ \psi ,\varphi_1^{0} \right]_{0+} \\
     W \left[ \psi ,\varphi_1^{0} \right]_{0-}  \\
  \end{array}
\right),
\hspace{10mm} 
B'_{0}(\psi) :=
\left(
  \begin{array}{r}
     W \left[ \psi ,\varphi_2^{0} \right]_{0+} \\
     -W \left[ \psi ,\varphi_2^{0} \right]_{0-} \\
  \end{array}
\right),
\label{bveczero}
\end{align}
where
$W[\psi_1,\psi_2]_{\phi\pm} = \lim_{\epsilon \to \pm 0} W \left[ \psi_1 , \psi_2 \right](\phi + \epsilon)$ are the limiting values of the Wronskian,
\begin{align}
W\left[ \psi_1 , \psi_2 \right] = 
\psi_1\frac{d\psi_2}{d\phi}-\psi_2\frac{d\psi_1}{d\phi}.
\end{align}
The functions $\varphi_i^{0}$ $(i=1,2)$ appearing in (\ref{bveczero})
are real eigenfunctions of $H_\Omega$ around $\phi = 0$, called the \lq reference modes\rq, which are  normalized with respect to the Wronskian, 
$W \left[ \varphi_1^{0},\varphi_2^{0} \right]=1$.
The most general form of the connection condition at $\phi =  0$, which ensures the local continuity of the probability current there, is then given by
\begin{align}
(U_0-{\bf 1}_2)B_0(\psi)+i(U_0+{\bf 1}_2)B_0'(\psi)=0,
\label{cconzero}
\end{align}
where  $U_0 \in U(2)$ is an arbitrary unitary matrix characterizing the connection condition (hence the singularity), and 
${\bf 1}_2$ is the identity matrix.  In components, the condition (\ref{cconzero}) consists of two equations linear in $\psi$ and its derivative $\frac{d\psi}{d\phi}$ at the singularity.
The use of the Wronskians is to render the equations well-defined, since the quantities 
$\psi$ and $\frac{d\psi}{d\phi}$ may be divergent at the singularity.

To provide the connection conditions at other singularities with $k = 1, \ldots, 5$ in (\ref{singpoints})  analogously to the case  $k = 0$,  we need to choose the reference modes $\varphi_i^k$  $(i=1,2)$ around each of the singularities $\phi_k$.   On account of the formal invariance of the Hamiltonian under $D_6$,  these reference modes may simply be provided by the translations,
\begin{align}
\varphi_i^k(\phi) := \varphi_i^0(\phi - \phi_k), 
\qquad k = 1, \ldots, 5.
\label{refmodes}
\end{align}
Note that, if we choose the reference modes for $k = 0$ possessing the parity property,
\begin{align}
\varphi_i^0(\phi) = (-1)^i \varphi_i^0(- \phi),
\end{align}
then (\ref{refmodes}) implies 
\begin{align}
\varphi_i^{k'} (\phi) = (-1)^i \varphi_i^k(R_l \phi),
\label{mirrorref}
\end{align}
which are fulfilled by the pair of 
reference modes at the two singular points $\phi_{k}$ and $\phi_k'$ mapped 
under the reflection $R_l$ as 
\begin{align}
\phi_{k'} = R_l \phi_k.
\label{mirref}
\end{align}
With this choice of the reference modes, following (\ref{bveczero}) 
we introduce the boundary vectors 
\begin{align}
B_{k}(\psi) :=
\left(
  \begin{array}{c}
     W \left[ \psi ,\varphi_1^{k} \right]_{\phi_k+} \\
     W \left[ \psi ,\varphi_1^{k} \right]_{\phi_k-}  \\
  \end{array}
\right),
\hspace{10mm} 
B'_{k}(\psi) :=
\left(
  \begin{array}{r}
     W \left[ \psi ,\varphi_2^{k} \right]_{\phi_k+} \\
     -W \left[ \psi ,\varphi_2^{k} \right]_{\phi_k-} \\
  \end{array}
\right),
\label{bveck1}
\end{align}
for $k = \hbox{even}$, and
\begin{align}
B_{k}(\psi) :=
\left(
  \begin{array}{c}
     W \left[ \psi ,\varphi_1^{k} \right]_{\phi_k-} \\
     W \left[ \psi ,\varphi_1^{k} \right]_{\phi_k+}  \\
  \end{array}
\right),
\hspace{10mm} 
B'_{k}(\psi) :=
\left(
  \begin{array}{r}
    - W \left[ \psi ,\varphi_2^{k} \right]_{\phi_k-} \\
     W \left[ \psi ,\varphi_2^{k} \right]_{\phi_k+} \\
  \end{array}
\right), 
\label{bveck2}
\end{align}
for $k = \hbox{odd}$.
As in (\ref{cconzero}), the connection conditions at $\phi = \phi_k$ for all $k$ are provided by
\begin{align}
(U_k-{\bf 1}_2)B_k(\psi)+i(U_k+{\bf 1}_2)B_k'(\psi)=0,
\label{cconk}
\end{align}
with matrices $U_k \in U(2)$ characterizing the singularities at $\phi = \phi_k$. 
We mention that (\ref{mirrorref}) implies  that  these boundary vectors are related as
\begin{align}
B_{k}(R_l\psi)   =   B_{k'}(\psi) , \qquad
B_{k}'(R_l\psi)   =   B_{k'}'(\psi),
\label{bmirror}
\end{align}
for the two singular points connected by (\ref{mirref}).  

Now, turning to the radial part, we introduce from the radial Hamiltonian 
$H_{r, \lambda}$ the new operator
\begin{align}
{\cal H}_{r,\lambda} := \sqrt{r} \circ H_{r,\lambda} \circ \frac{1}{\sqrt{r}} =
- \frac{d^2}{d   r^2} +\frac{3}{8} \omega^2 r^2 +
\frac{\lambda-\frac{1}{4}}{r^2}.
\end{align}
Note that the self-adjointness of ${\cal H}_{r, \lambda}$ with respect to the measure $dr$ implies
the self-adjointness of $H_{r, \lambda}$ with respect to the radial measure $rdr$.   
It is known (see, {\it e.g.},  \cite{Tsutsui2}) that for $\lambda \ge 1$ the operator ${\cal H}_{r, \lambda}$ is essentially self-adjoint, while for $\lambda < 1$  it admits a $U(1)$ family of self-adjoint extensions.  The family can be specified by the boundary condition for the radial wave function $\rho(r)$, which is given, using the Wronskian in the radial part, by
\begin{align}
\frac{W \left[ \rho , \varphi_{1} \right]_{0+}}{W \left[ \rho , \varphi_{2} \right]_{0+}}
=-\kappa(\lambda).
\label{radial boundary condition}
\end{align}
Here, $\varphi_{i} (r)$ $(i=1,2)$ are the reference modes for the radial part which are eigenfunctions of ${\cal H}_{r,\lambda}$ and normalized as 
$W \left[ \varphi _{1} ,\varphi_{2}  \right]=1$.  
Note that the real number $\kappa(\lambda)$ (which includes $\kappa(\lambda) = \infty$) characterizes the singularity at $r=0$ and is dependent on the angular eigenvalue $\lambda$ in general.

\subsection{Mirror-$S_3$ and scale invariant quantizations}

In our scheme of quantization, we have the set of parameters in $U_k$, $k = 0, \ldots, 5$ and
$\kappa(\lambda)$ which specify the connection/boundary conditions and thereby the self-adjoint extensions of the Hamiltonian.
These conditions are subjected to the symmetry we wish to bestow with the quantized model.   
{}For the symmetry of the model, one may assume, for instance, that the three particles are identical physically.   
Since the three possible exchanges of the particles are generated by the reflections, $P_i$ ($i = 1, 2, 3$),  which form an $S_3$ group, the invariance under the reflections implies 
the \lq exchange-$S_3$\rq\  group as the symmetry of the model.  
One may further assume that the pairwise collision be physically independent on the position of the remaining spectator particle.  
This leads to the $D_6$ symmetry mentioned in the Introduction, and the quantizations with this symmetry has  been presented in \cite{Tsutsui2}.   

In this paper, we relax our demand and assume only the symmetry under the  \lq mirror-$S_3$'
group, which is generated by the reflections, $R_i$ ($i = 1, 2, 3$), consisting of the elements
$\{e, R_i, (\mathcal{R}_{\frac{\pi}{3}})^{\pm 2}\}$ with $e$ being the identity.
In physical terms, this demand ensures the invariance of the model under all cyclic permutations of the particles; for example, 
if $x_1<x_2<x_3$, the interaction in the limit $x_2 \to x_1$ is identical to that in the limit $x_2 \to x_3$
when the spectator particle is fixed.  
In our scheme, this is equivalent to the requirement that all $U_k$ be identical, {\it i.e.},
\begin{align}
    U_{k}   =   U,
    \hspace{10mm}
k= 0, 1,2,3,4,5,
\label{umirror}
\end{align}
for some $U \in SU(2)$.   Indeed, when combined with (\ref{bmirror}), the property (\ref{umirror}) ensures 
that the boundary conditions are compatible with all the reflections in the mirror-$S_3$ group.
Once the mirror-$S_3$ is installed as a symmetry in the angular part, we can classify the eigenstates of the operator $H_{\Omega}$ in (\ref{angular equation})
in terms of the representations of the group $S_3$, which has  two 
1-dimensional and one 2-dimensional irreducible representations (see Table \ref{table of ms3}).


\begin{table}[t]
 \begin{center}
 {\arrayrulewidth=0.2pt
  \begin{tabular}{||||rl||||c|c|c||||}
    \hline\hline\hline\hline
     \multicolumn{2}{||||c||||}{conjugacy class}& $\{e\}$   & $\{R_i\}$   & $\{(\mathcal{R}_{\frac{\pi}{3}})^{\pm 2} \}$  \\
    \hline\hline\hline\hline
       identity rep.&$\chi^+$ &  $1$  &  $1$  &  $1$  \\
    \hline
       signature rep.&$\chi^-$ &  $1 $ &  $-1 $ &  $1 $  \\
    \hline
       2-dim. rep.&$\chi^{(2)}$ & $ 2$  &  $0 $ &  $-1$  \\
    \hline\hline\hline\hline
  \end{tabular}}
 \end{center}
    \caption{Character table of mirror-$S_3$}
\label{table of ms3}
\end{table}

Another property we wish to have comes from the observation that the Hamiltonian (\ref{threeham}) acquires a formal scale invariance
in the limit $\omega \to 0$.   Namely, we demand that this scale symmetry be maintained in the limit in our inequivalent quantizations, so that our quantized model is smoothly connected to the model with a pure inverse square potential.
Roughly speaking, the scale symmetry breaks down at the quantum level when we allow the parameters of self-adjoint extensions to bear nontrivial scale dimensions (see \cite{Tsutsui1}).    For the mirror-$S_3$ invariant 
quantizations we are considering, a sufficient condition prohibiting such scale parameters is that 
\begin{align}
U = \pm {\bf 1}_2,  
 \hspace{10mm}
 \hbox{or} 
 \hspace{10mm}
U = V \sigma_3 V^{-1}, 
\hspace{5mm}
V \in SU(2),
\label{scale invariant matrices}
\end{align}
for the angular part  specified by (\ref{umirror}), and that
\begin{align}
\kappa(\lambda) = 0,
\qquad
 \hbox{or} 
\qquad
\kappa(\lambda) = \infty, 
\qquad \hbox{for each}  \,\,\, \lambda,
\label{radial boundary condition2}
\end{align}
for the radial part (\ref{radial boundary condition}).  In the present paper, we shall restrict ourselves to the class of inequivalent  quantizations fulfilling (\ref{umirror}), (\ref{scale invariant matrices}) and (\ref{radial boundary condition2}).

\section{Angular part}

We now discuss inequivalent quantizations for the angular part in detail.  
Among the scale invariant choices (\ref{scale invariant matrices}) the cases $U = \pm {\bf 1}_2$, 
which are parity invariant, have already been treated in \cite{Tsutsui2} and will not be discussed here.
To analyze the remaining cases $U=V \sigma_3 V^{-1}$, we consider $V$ for which 
$U \neq \pm \sigma_3$ and $U = \pm \sigma_3$, separately.  
This separate treatment is required since for the latter case the angular part breaks into six sectors 
$s_k = (\phi_{k-1}, \phi_k)$, $k = 1, \dots, 6$, which are physically disconnected at the singular points (see  Fig.\ref{xy-plane}).  
The treatment is based on a set of basic solutions, which are used to analyze both the connected and the separated cases later.  

\subsection{Basic solutions}

{}For convenience, 
we first set $\lambda=9\mu^2$ allowing for complex $\mu$ for $\lambda < 0$.
The two independent solutions for (\ref{radial equation}) are given \cite{Tsutsui2} by 
\begin{align}
       v_{1,\mu}(\phi)=& |\sin 3 \phi|^\nu
       \ F\left(\frac{\nu+\mu}{2},\frac{\nu-\mu}{2},\nu+\frac{1}{2};\sin ^2 3 \phi\right),  \\
       v_{2,\mu}(\phi)=& |\sin 3 \phi|^{1-\nu}
       \ F\left(\frac{1-\nu-\mu}{2},\frac{1-\nu+\mu}{2},\frac{3}{2}-\nu;\sin ^2 3 \phi\right),
\end{align}
where $F(\alpha,\beta,\gamma;z)$ is the standard hypergeometric function.
{}For the reference modes needed in the connection condition 
at the singularity $\phi = \phi_0$, we choose
\begin{align}
\varphi^0_1 (\phi) &= {1\over{\sqrt{3(2\nu-1)}}}  v_{1,\mu_0}(\phi) \left[ \Theta(\phi)-\Theta(-\phi) \right],
\\
\varphi^0_2 (\phi) &= - {1\over{\sqrt{3(2\nu-1)}}}  v_{2,\mu_0}(\phi),
\end{align}
with some real $\mu_0$, where $\Theta(\phi)$ is the Heaviside step function.
The reference modes for the connection condition 
at the singularity $\phi = \phi_k$ are provided according to (\ref{refmodes}). 
We introduce the shorthand,
\begin{align}
a_1(\mu)&=v_{1,\mu}\left(\frac{\pi }{6}-0\right)=
\frac{\Gamma (\nu +\frac{1}{2})\Gamma (\frac{1}{2})}{\Gamma (\frac{\nu +\mu+1 
}{2})\Gamma (\frac{\nu - \mu+1 }{2})},\\
a_2(\mu)&=v_{2,\mu}\left(\frac{\pi }{6}-0\right)=
\frac{\Gamma (\frac{3}{2}-\nu) \Gamma (\frac{1}{2})}{\Gamma (1-\frac{\nu +\mu}
{2})\Gamma (1-\frac{\nu - \mu }{2})},\\
b_1(\mu)&=v'_{1,\mu}\left(\frac{\pi }{6}-0\right)=
\frac{6 \  \Gamma (\nu +\frac{1}{2})\Gamma (\frac{1}{2})}{\Gamma (\frac{\nu +\mu }
{2})\Gamma (\frac{\nu - \mu }{2})},\\
b_2(\mu)&=v'_{2,\mu}\left(\frac{\pi }{6}-0\right)=
\frac{6 \  \Gamma (\frac{3}{2}-\nu) \Gamma (\frac{1}{2})}{\Gamma (\frac{1-\nu - \mu}{2})\Gamma (\frac{1-\nu + \mu }{2})}.
\end{align}
and 
\begin{align}
\alpha=\frac{a_1 a_2}{3(2\nu-1)},\hspace{10mm}
\beta=\frac{b_1 b_2}{3(2\nu-1)},\hspace{10mm}
\gamma=\frac{a_1 b_2+a_2 b_1}{3(2\nu-1)}
=-\frac{\cos \pi \mu}{\cos \pi \nu },
\label{relgamma}
\end{align}
which fulfill the relation
\begin{align}
\alpha \beta=\frac{1}{4}(\gamma^2-1).
\label{the relation of abc}
\end{align}

In terms of these, as a set of basic solutions in sector  $s_1$,
we furnish the symmetric and anti-symmetric eigenfunctions,
\begin{align}
\eta^1_{+,\mu}(\phi)&=
\left\{
  \begin{array}{ll}
    b_2(\mu)v_{1,\mu}(\phi) -b_1(\mu)v_{2,\mu}(\phi)   &
    (0<\phi\le\frac{\pi}{6})    \\
    b_2(\mu)v_{1,\mu}(\frac{\pi}{3}-\phi) -b_1(\mu)v_{2,\mu}(\frac{\pi}{3}-\phi)   &
    (\frac{\pi}{6}\le \phi<\frac{\pi}{3})    \\
    0   &    (2 \pi>\phi> \frac{\pi}{3}) \\
  \end{array}
\right.\\
\eta^1_{-,\mu}(\phi)&=
\left\{
  \begin{array}{ll}
    a_2(\mu)v_{1,\mu}(\phi) -a_1(\mu)v_{2,\mu}(\phi)   &
    (0<\phi\le\frac{\pi}{6})    \\
    -a_2(\mu)v_{1,\mu}(\frac{\pi}{3}-\phi) +a_1(\mu)v_{2,\mu}(\frac{\pi}{3}-\phi)   &
    (\frac{\pi}{6}\le \phi<\frac{\pi}{3})    \\
    0   &    (2 \pi>\phi> \frac{\pi}{3}) \\
  \end{array}
\right.
\end{align}
Likewise, we also introduce the basic solutions in sector $s_k$ from these by translation,
\begin{align}
\eta^k_{\pm,\mu}(\phi) =\eta^1_{\pm,\mu}\left(\phi-\phi_{k-1}\right), 
\hspace{10mm} k=2, \ldots, 6.
\end{align}
The general solution is then given by a linear combination of these basic solutions with appropriate coefficients $c_{\pm}^k$,
\begin{align}
\psi_\mu(\phi) =\sum_{k=1}^6 \left(c_+ ^k \eta ^k _{+,\mu}(\phi)+c_-^k \eta ^k _{-,\mu}(\phi)\right).
\label{generalsol}
\end{align}
Out of these basic solutions each having their support in one sector, we also introduce a set of
solutions having supports on the sectors $s_k$ with $k$ odd or even only:
\begin{align}
\eta_+& =\sum_{n=1}^3 \eta^{2n-1}_{+,\mu},&
\tilde{\eta}_+& =\sum_{n=1}^3 \eta^{2n}_{+,\mu},\\
\eta_-& =\sum_{n=1}^3 (-1)^n \eta^{2n-1}_{-,\mu},&
\tilde{\eta}_-& =\sum_{n=1}^3 (-1)^n \eta^{2n}_{-,\mu}.
\end{align}

\subsection{Connected case}

To analyze the connected case 
\begin{align}
U=V \sigma_3 V^{-1}, \qquad 
U \neq \pm \sigma_3,
\end{align}
we parametrize the characteristic matrix, using a real $\xi$ and a complex $\zeta$, as
\begin{align}
U=\left(
  \begin{array}{cc}
    \xi & \zeta \\
    \zeta^*  & -\xi \\
  \end{array}
\right),
\hspace{10mm}
\xi^2 + \vert \zeta \vert^2 = 1.
\label{umat}
\end{align}
In terms of the vectors defined from the coefficients in the general solutions (\ref{generalsol}),
\begin{align}
C^k = \left(
  \begin{array}{c}
    c^k_+   \\
    c^k_-   \\
  \end{array}
\right),
\end{align}
we find that the connection conditions (\ref{cconk}) become the matrix equations,
\begin{align}
C^{2n}
=T_+ C^{2n-1},
\hspace{10mm}
C^{2n+1}
=T_- C^{2n},
\end{align}
with the transfer matrices defined at the singularities between odd-even and even-odd sectors,
\begin{align}
T_+=
\frac{1}{\zeta}
\left(
  \begin{array}{cc}
    -\xi+\gamma & -2\alpha  \\
    -2\beta & \xi+\gamma  \\
  \end{array}
\right),
\hspace{10mm}
T_-=
\frac{1}{\zeta^*}
\left(
  \begin{array}{cc}
    \xi+\gamma & -2\alpha  \\
    -2\beta & -\xi+\gamma  \\
  \end{array}
\right),
\end{align}
where $\alpha$, $\beta$ and $\gamma$ are given in (\ref{relgamma}).  
These relations lead to the consistency condition,
\begin{align}
T C^1=C^1, \qquad 
T :=T_-T_+T_-T_+T_-T_+.
\end{align}
{}From this we find
$\textrm{det}(T-I)=0$, which implies
\begin{align}
\gamma^2 =1,
\qquad \hbox{or} \qquad
\gamma^2 =\frac{1+3\xi^2}{4}.
\label{the case of 1,2-dim}
\end{align}
These determine the spectrum $\mu$ in the angular part.

To examine the case $\gamma^2 =1$  in  (\ref{the case of 1,2-dim}), we
note from (\ref{the relation of abc}) that the condition 
implies that either $a_1$, $ a_2$, $ b_1$ or $ b_2$ must vanish.   
None of these are compatible with each other, and we 
consider them separately.  First, we find that $a_1=0$ occurs when
\begin{align}
\mu=\nu+1+2m, \qquad
m = 0, 1, \ldots.
\label{spca1}
\end{align}
We then have
\begin{align}
T =\left(
\begin{array}{cl}
    1   &  0  \\
    \frac{12\beta}{1-\xi}   &  1  \\
  \end{array}
\right),
\end{align}
and from this we find the coefficient vectors $C^k$ to obtain, 
up to a constant,  
the corresponding eigenfunctions
\begin{align}
\tilde{\psi}_{-}(\phi) =\frac{1-\xi}{\zeta^*}\eta_-(\phi) + \tilde{\eta}_- (\phi).
\end{align}
These eigenfunctions belong to the signature representation $\chi^-$ of $S_3$.
Second, $a_2=0$ occurs when
\begin{align}
\mu=2m+2-\nu, \qquad
m = 0, 1, \ldots
\label{spca2}
\end{align}
and the transfer matrix is
\begin{align}
T =\left(
  \begin{array}{cl}
    1   &  0  \\
    -\frac{12\beta}{1+\xi}   &  1  \\
  \end{array}
\right).
\end{align}%
The eigenfunctions are found to be
\begin{align}
\psi_{-} (\phi)=\eta_- (\phi)- \frac{1-\xi}{\zeta} \tilde{\eta}_- (\phi),
\end{align}
which again belong to the signature representation $\chi^-$ of $S_3$.
Third,  $b_1=0$ occurs when
\begin{align}
\mu=\nu+2m, \qquad
m = 0, 1, \ldots.
\label{spcb1}
\end{align}
We then have
\begin{align}
T =\left(
  \begin{array}{cl}
    1   &   \frac{12\alpha}{1+\xi}  \\
    0  &  1  \\
  \end{array}
  \right),
\end{align}
and the eigenfunctions
\begin{align}
\tilde{\psi}_{+} (\phi)= - \frac{1-\xi}{\zeta^*} \eta_+ (\phi) + \tilde{\eta}_+ (\phi),
\end{align}
which belong to the identity representation $\chi^+$ of $S_3$.
Finally, if $b_2=0$, we find
\begin{align}
\mu=|1-\nu+2m|, \qquad
m = 0, 1, \ldots
\label{spcb2}
\end{align}
and 
\begin{align}
T =\left(
  \begin{array}{cl}
    1   &   -\frac{12\alpha}{1-\xi}  \\
    0  &  1  \\
  \end{array}
  \right).
\end{align}
We obtain the eigenfunctions
\begin{align}
\psi_{+} (\phi) =\eta_+ (\phi) + \frac{1-\xi}{\zeta} \tilde{\eta}_+ (\phi),
\end{align}
belonging to the identity representation $\chi^+$ of $S_3$.

\vspace{1cm}
\begin{figure}[t]
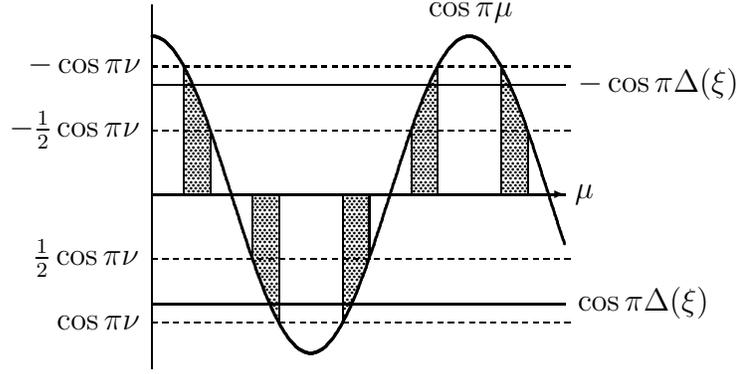

  \begin{center}
%
  \end{center}
  \caption{%
Eigenvalues $\mu$ arise at the intersections of  $\cos \pi \mu$ and the two lines $\pm \cos\Delta(\xi)$, which are shown here for  the values $\nu=0.8$ and $\xi=\frac{5}{6}$.
Dashed lines represent the upper and the lower limits of the lines, and the shaded regions indicate the allowed ranges of the eigenvalues. }
\label{cos pi mu}
\end{figure}

When the second condition $\gamma^2 =\frac{1+3\xi^2}{4}$ 
in (\ref{the case of 1,2-dim}) holds, on the other hand,
we have $T= {\bf 1}_2$ and hence $C^1$ is left undetermined.  For definiteness we 
may choose $C^1$ as an eigenvector of the rotation $\mathcal{R}_\frac{\pi}{3}^2$.
The transfer matrix that corresponds to $\mathcal{R}_\frac{\pi}{3}^2$ is
\begin{align}
T_-T_+=\left(
  \begin{array}{cc}
    -\frac{1}{2}   &  -\frac{4\alpha(\gamma+\xi)}{1-\xi^2}  \\
    -\frac{4\beta(\gamma-\xi)}{1-\xi^2}   &  -\frac{1}{2}  \\
  \end{array}
\right),
\end{align}
which has
the eigenvectors 
\begin{align}
X_{\pm} =\left(
  \begin{array}{c}
    2\alpha   \\
    \pm\sqrt{3}i(-\xi+\gamma)   \\
  \end{array}
\right).
\end{align}
With these eigenvectors, the vectors $C_\pm^k$ may be written as
\begin{align}
C_\pm^{2n-1}=\jmath^{\mp n} X_\pm,
\qquad
C_\pm^{2n}=\jmath^{\mp n} T_+ X_\pm,
\qquad
\jmath :=\frac{-1 + i\sqrt{3}}{2}.
\end{align}
If we denote by
$\psi_1^{(\pm)}$ the eigenfunction corresponding to $C_\pm^k$ with $\gamma > 0$, 
and similarly by $\psi_2^{(\pm)}$ the eigenfunction corresponding to  $C_\pm^k$ with $\gamma < 0$, then 
\begin{align}
\mathcal{R}^2_\frac{\pi}{3}\psi_i ^{(\pm)} = \jmath^{\pm 1} \psi_i^{(\pm)},
\qquad
R_2 \psi_i ^{(\pm)} &= \psi_i^{(\mp)},\qquad i = 1, 2 .
\end{align}

\bigskip
\begin{table}[t]
\begin{center}
{\arrayrulewidth =0.2pt
\renewcommand\arraystretch{1.4}
\begin{tabular}{||||c||||c|c||||}
    \hline\hline \hline\hline
    eigenvalue $\mu$ &eigenfunction& rep. of $S_3$ \\
    \hline \hline \hline\hline
    $2m+1+\nu$&
    $\tilde{\psi}_{-}$&
    $\chi^-$  \\
    \hline
    $2m+2-\nu$& 
    $\psi_{-}$&
    $\chi^-$  \\
    \hline
    $2m+\nu$&
    $\tilde{\psi}_{+}$&
    $\chi^+$ \\
    \hline
    $|2m+1-\nu|$&
    $\psi_{+}$&
    $\chi^+$ \\
    \hline
    $2m+1-\Delta(\xi)$&&\\
    $2m+1+\Delta(\xi)$&
    \raisebox{13pt}[0pt][0pt]{\parbox{45pt}{$ \psi_1 ^{(+)} ,\  \psi_1 ^{(-)}  $}}&
    \raisebox{13pt}[0pt][0pt]{\parbox{15pt}{$\chi^{(2)}$}}  \\
    \hline
    $2m+2-\Delta(\xi)$&&\\
    $2m+\Delta(\xi)$&   
    \raisebox{13pt}[0pt][0pt]{\parbox{45pt}{$ \psi_2 ^{(+)} ,\  \psi_2 ^{(-)}  $}}&
    \raisebox{13pt}[0pt][0pt]{\parbox{15pt}{$\chi^{(2)}$}} \\
    \hline\hline \hline\hline
    \end{tabular}}
\end{center}
\caption{Eigenvalues and eigenfunctions in the connected case.}
\label{casen3-2t}
\end{table}

To find the spectrum of these eigenstates, we combine  (\ref{the case of 1,2-dim}) and the last equation of  (\ref{relgamma}) to obtain
\begin{eqnarray}
\cos \pi \mu = \mp \cos \pi \Delta(\xi),
\qquad
\Delta (\xi) := \frac{1}{\pi} \textrm{Arccos} 
    \left( \frac{\sqrt{1+3\xi^2}}{2} \cos \pi \nu \right).
\end{eqnarray}
The solutions for $\mu$ will furnish the eigenvalues, which occur periodically as displayed 
in Fig.\ref{cos pi mu}.
We then see that the eigenvalues for $\gamma > 0$ are given by
\begin{eqnarray}
\mu = \left\{
  \begin{array}{l}
   2m +1- \Delta(\xi) , \\
   2m+1+\Delta(\xi),  \\
  \end{array}
\right.
\end{eqnarray}
whereas 
the eigenvalues for $\gamma < 0$ are
\begin{eqnarray}
\mu = \left\{
  \begin{array}{l}
    2m+2 -\Delta(\xi) , \\
    2m+\Delta (\xi) ,   \\
  \end{array}
\right.
\end{eqnarray}
where $m = 0, 1, \ldots $ .
Since $\vert \xi \vert \le 1$, we find that $\Delta(\xi)$ stays in the range,
\begin{align}
 \Delta (0)  \le  \Delta(\xi)  \le \Delta (1) = \nu.
\end{align}
The eigenvalues and eigenfunctions are summarized in Table \ref{casen3-2t}.

\subsection{Separated case}

Next we analyze the case 
\begin{align}
U = \sigma_3,
\label{sigma3}
\end{align}
under which all the sectors $s_k$ are physically separated from each other.   (The other separated case $U = -\sigma_3$ can be dealt with similarly and will not be discussed here.)
To proceed, we note that under the choice (\ref{sigma3})
the boundary condition (\ref{cconk}) reads 
\begin{align}
b_2 C_+^{2n-1} = a_2 C_-^{2n-1} = b_1 C_+^{2n} = a_1 C_-^{2n} =0.
\end{align}
As before, the vanishing conditions of $a_1$, $a_2$, $b_1$ and $b_2$ are not compatible, and we consider them separately.   
Recall, first, that $a_1=0$ is realized by $\mu$ in  (\ref{spca1}).  The vectors $C^k$ obtained in this case lead to the eigenfunctions $\tilde{\eta}_{- }$ belonging to the signature  representation $\chi^-$, and also
\begin{align}
\tilde{\eta}_{- }^{(\pm)} =\sum_{n=1}^3 \jmath^{\mp n} \eta^{2n}_{-,\mu},
\end{align}
which belong to the doublet representation $\chi^{(2)}$.
Second,  $a_2=0$ is realized by $\mu$ in  (\ref{spca2}), and the corresponding 
eigenfunctions are given by  $\eta_{- }$ belonging to $\chi^-$, and 
\begin{align}
\eta_{- }^{(\pm)} =\sum_{n=1}^3 \jmath^{\mp n} \eta^{2n-1}_{-,\mu},
\end{align}
which
belong to $\chi^{(2)}$.
Third, $b_1=0$ occurs when $\mu$ is given by  (\ref{spcb1}). 
The eigenfunctions are found to be $\tilde{\eta}_{+ }$ belonging to 
$\chi^+$, and
\begin{align}
\tilde{\eta}_{+ }^{(\pm)} =\sum_{n=1}^3  \jmath^{\mp n}  \eta^{2n}_{+,\mu},
\end{align}
which belong to $\chi^{(2)}$.
Finally, $b_2=0$ occurs when $\mu$ is given by  (\ref{spcb2}).
The eigenfunctions are $\eta_{+ }$  belonging to $\chi^+$, and 
\begin{align}
\eta_{+ }^{(\pm)} =\sum_{n=1}^3 \jmath^{\mp n} \eta^{2n-1}_{+,\mu},
\end{align}
which belong to $\chi^{(2)}$.
These eigenvalues and the eigenfunctions are summarized in Table \ref{casen3-1t}.  
The spectral behavior of the two cases
is shown in Fig.\ref{spectrum} (left).

\begin{table}[t]
\begin{center}
{\arrayrulewidth =0.2pt
\begin{tabular}{||||c||||c|c||||}
    \hline\hline\hline\hline
    eigenvalue $\mu$ & eigenfunction & rep. of $S_3$ \\
    \hline\hline\hline\hline
     &
    $\tilde{\eta}_{- }$ &
    $\chi^-$ \\
    \cline{2-3}
    \raisebox{11pt}[0pt][0pt]{\parbox{65pt}{$2m+1+\nu$}}        &
    $\tilde{\eta}^{(+)}_{- } \ ,\  \tilde{\eta}^{(-)}_{- }$ &
    $\chi^{(2)}$ \\
    \hline
    &
    $\tilde{\eta}_{+ }$ &
    $\chi^+$ \\
    \cline{2-3}
    \raisebox{11pt}[0pt][0pt]{\parbox{65pt}{$2m+\nu$}}&  
    $\tilde{\eta}^{(+)}_{+ } \ ,\  \tilde{\eta}^{(-)}_{+ }$ &
    $\chi^{(2)}$\\
    \hline   
    &  
    $\eta_{- }$ &
    $\chi^-$ \\
    \cline{2-3}
    \raisebox{11pt}[0pt][0pt]{\parbox{65pt}{$2m+2-\nu$}}&  
    $\eta^{(+)}_{- } \ ,\  \eta^{(-)}_{- }$ &
    $\chi^{(2)}$ \\
    \hline
    &
    $\eta_{+ }$ &
    $\chi^+$ \\
    \cline{2-3}
    \raisebox{11pt}[0pt][0pt]{\parbox{65pt}{$|2m+1-\nu|$}}& 
    $\eta^{(+)}_{+ } \ ,\  \eta^{(-)}_{+ }$ &
    $\chi^{(2)}$ \\
    \hline\hline\hline\hline
    \end{tabular}}
\end{center}
\caption{Eigenvalues and eigenfunctions in the separated case.}
\label{casen3-1t}
\end{table}

At this point, we mention that the connected case discussed earlier has a smooth limit to the separated case, that is,  
the eigenfunctions in the former case can be obtained formally from those in the latter case by 
considering the limit $U \to \sigma_3$,
even though the two cases require distinctive treatments.  
{}For the eigenstates which are singlets of the mirror-$S_3$, this can be seen at once since the eigenfunctions in the connected case reduce to 
\begin{align}
\psi_{+} \to \eta_{+ }, \qquad
\psi_{-} \to \eta_{- }, \qquad
\tilde{\psi}_{+} \to \tilde{\eta}_{+ } \qquad,
\tilde{\psi}_{-} \to \tilde{\eta}_{- },
\end{align}
in the limit $\xi \to 1$, which is equivalent to $U \to \sigma_3$; see (\ref{umat}).
To see that the same is true for the doublets,  consider 
the case $a_1=0$ in which the eigenstates vanish $\psi^{(\pm)}_+ \to 0$ in the limit.  Non-vanishing outcomes in the limit may be obtained, however, by
rescaling them properly as
\begin{align}
\frac{-8i}{\zeta^*\sqrt{3}}\, \psi^{(\pm)}_1 \to \tilde{\eta}^{(\pm)}_{- }.
\end{align}
Similarly, for $a_2=0$, we find 
\begin{align}
\frac{i}{2\sqrt{3}} \psi^{(\pm)}_2 &\to \eta^{(\pm)}_{- }.
\end{align}
The continuity in the remaining cases $b_1=0$, $b_2=0$ can also be argued analogously.

\section{Radial part and the total energy spectrum}

 \subsection{Radial part}

The eigenfunctions for the radial part can be obtained immediately, since there remain only two choices (\ref{radial boundary condition2}) for the inequivalent quantizations under the scale invariance we have imposed.  Note that the radial Hamiltonian  $\mathcal{H
}_{r,\lambda}$ admits 
the two independent solutions \cite{Tsutsui2},
\begin{align}
\rho_{E,1} (r)&=(\sqrt{c}r)^{\sqrt{\lambda}+\frac{1}{2}} 
    \ e^{-\frac{1}{2}c r^2}
    \ \Phi \left(\frac{1+\sqrt{\lambda}}{2}  -\frac{E}{4c} ,1+\sqrt{\lambda};c r^2\right),\\
\rho_{E,2} (r)&=(\sqrt{c}r)^{-\sqrt{\lambda}+\frac{1}{2}} 
    \ e^{-\frac{1}{2}c r^2}
    \ \Phi \left(\frac{1-\sqrt{\lambda}}{2}  -\frac{E}{4c} ,1-\sqrt{\lambda};c r^2\right),
\end{align}
where $\Phi(a,b;z)$ is the confluent hypergeometric function
and $c:=\sqrt{\frac{3}{8}}\omega$.
Since these two solutions diverge generically as $r \to \infty $, 
the square integrability requires that the eigenfunctions be proportional to their linear combination,
\begin{align}
\rho_{E}(r)=
\frac{\Gamma(1-\sqrt{\lambda})}{\Gamma(\frac{1-\sqrt{\lambda}}{2} -\frac{E}{4c})} \rho_{E,1}(r)
-\frac{\Gamma(1+\sqrt{\lambda})}{\Gamma(\frac{1+\sqrt{\lambda}}{2} -\frac{E}{4c})} \rho_{E,2}(r),
\end{align}
which vanishes as $r \to \infty $.

To enforce (\ref{radial boundary condition2}),  we choose the reference modes as
\begin{align}
\varphi_{1}(r)&=\left(2\sqrt{\lambda c}\right)^{-1/2} \rho_{E,1}(r),&
\varphi_{2}(r)&=-\left(2\sqrt{\lambda c}\right)^{-1/2} \rho_{E,2}(r).
\end{align}
With these the radial boundary condition (\ref{radial boundary condition}) reads
\begin{align}
\kappa(\lambda) =
\frac{\Gamma(1+\sqrt{\lambda})\, \Gamma(\frac{1-\sqrt{\lambda}}{2}-\frac{E}{4c})}{\Gamma(1-\sqrt{\lambda})\, 
\Gamma(\frac{1+\sqrt{\lambda}}{2}-\frac{E}{4c})}.
\label{kappa value}
\end{align}
{}For the case $\kappa(\lambda) = 0$, the solutions and the energy eigenvalues are
\begin{align}
\rho_{E}(r) = r^{\sqrt{\lambda} + {1\over 2}}e^{-\frac{1}{2}cr^2}L^{\sqrt{\lambda}}_m(cr^2),
\qquad
E= E(m, \lambda) = 2c\left(2m+1+\sqrt{\lambda}\right),
\label{radial solution1}
\end{align}
for $m = 0, 1, \ldots$, where $L^{\sqrt{\lambda}}_m$ is the (generalized) Laguerre polynomial.
Analogously, for the case $\kappa(\lambda) = \infty$
we find
\begin{align}
\rho_{E}(r) = r^{- \sqrt{\lambda} + {1\over 2}}e^{-\frac{1}{2}cr^2}L^{-\sqrt{\lambda}}_m(cr^2),
\qquad 
E = E(m, \lambda) = 2c\left(2m+1-\sqrt{\lambda}\right),
\label{radial solution2}
\end{align}
for $m = 0, 1, \ldots$.    As noted earlier, these two types of solutions arise only for $\lambda < 1$;
otherwise only the former solution (\ref{radial solution1}) is allowed.  
This solution (\ref{radial solution1})  is  in fact the one 
used by Calogero \cite{Calogero1} and also conventionally adopted by others \cite{Gibbons-Townsend,Polychronakos2,Markvardsen-Johnson} for all values of $\lambda$.

\subsection{Total energy spectra}

Having obtained the solutions for both the radial and the angular parts, we now
construct the solutions for the entire (relative coordinates) system (\ref{angular equation}) by combining the solutions of the respective parts as
\begin{align}
\psi_{E} (r;\lambda) = {1\over{\sqrt{r}}} \rho_E(r)\, \psi_\mu(\phi), \qquad
E = E(m, \lambda = 9\mu^2),
\end{align}
where we have denoted the angular solutions by $\psi_\mu(\phi)$ collectively.
These solutions $\rho_E(r)$ and $\psi_\mu(\phi)$ are dependent on the parameters 
$\kappa(\lambda)$ and $\xi, \zeta$ that specify the inequivalent quantizations of the respective part.   

It is important, however, to note that not all combinations of them are allowed
because the choice $\kappa(\lambda) = \infty$ is available only 
if $\lambda < 1$.    In fact, one can readily see that $\kappa(\lambda) = \infty$ is possible, at most,  for the lowest
two angular eigenvalues $\lambda = \lambda_1$ or $\lambda = \lambda_2$, where
\begin{align}
\lambda_1 := 9 (1-\nu)^2,
\qquad
\lambda_2 := 9 (1-\Delta)^2,
\end{align}
and that $\kappa(\lambda) = 0$ for all other $\lambda$ which are above the two.
More explicitly, the case $\kappa(\lambda) = \infty$ is admitted for $\lambda_1$ if
\begin{align}
\frac{2}{3}<\nu<\frac{4}{3},
\label{nucon1}
\end{align}
and for $\lambda_2$ if
\begin{align}
1-\nu_0<\nu<1+\nu_0,
\hspace{10mm}
\nu_0= \frac{1}{\pi} \left\vert \arctan \sqrt{3} \xi \right\vert.
\label{nucon2}
\end{align}
Since the condition (\ref{nucon2}) is stricter than (\ref{nucon1}),  available choices  for the inequivalent quantizations depend on the value of $\nu$, as summarized in Table \ref{kappa_mu}.

\begin{table}[t]
\begin{center}
{\arrayrulewidth =0.2pt
\begin{tabular}{||||c||||c|c||||}
    \hline\hline\hline\hline
    range of $\nu$ & $\kappa(\lambda_1)$ & $\kappa(\lambda_2)$ \\
    \hline\hline\hline\hline
    $\frac{1}{2}<\nu\le\frac{2}{3}$ & $0$  & $0$ \\
    \hline
    $\frac{2}{3}<\nu\le 1-\nu_0$ & $0$ or $\infty$ & $0$ \\
    \hline
    $1-\nu_0 < \nu < 1+\nu_0$ & $0$ or $\infty$ & $0$ or $\infty$\\
    \hline
    $1+\nu_0 \le \nu < \frac{4}{3}$ & $0$ or $\infty$ & $0$ \\
    \hline
    $\frac{4}{3} \le \nu < \frac{3}{2}$ & $0$ & $0$ \\
    \hline\hline\hline\hline
\end{tabular}}
\end{center}
\caption{Possible choices of $\kappa(\lambda_1)$ and $\kappa(\lambda_2)$.}
\label{kappa_mu}
\end{table}

\begin{figure}[hp]
  \begin{center}
\begin{picture}(400,460)(20,-10)
{ 

\put(0,0){\vector(0,1){450}}
\put(0,0){\vector(1,0){200}}

\put(-3,457.5){$\mu$}
\put(206,-2.5){$\xi$}

\put(-3,100){\line(1,0){6}}
\put(-3,200){\line(1,0){6}}
\put(-3,300){\line(1,0){6}}
\put(-3,400){\line(1,0){6}}

\put(40,-3){\line(0,1){6}}
\put(80,-3){\line(0,1){6}}
\put(120,-3){\line(0,1){6}}
\put(160,-3){\line(0,1){6}}

\put(-10,-2,5){$0$}
\put(-10,97.5){$1$}
\put(-10,197.5){$2$}
\put(-10,297.5){$3$}
\put(-10,397.5){$4$}

\put(37.5,-12.5){$0$}
\put(77,-12.5){$\frac{1}{3}$}
\put(117,-12.5){$\frac{2}{3}$}
\put(157.5,-12.5){$1$}

\put(35.3,7){$\odot^1$}
\put(35.3,105.7){$\boxdot^1$}
\put(35.3,207){$\odot^1$}
\put(35.3,305.7){$\boxdot^1$}
\put(35.3,407){$\odot^1$}
\put(35.3,87){$\odot^1$}
\put(35.3,185.7){$\boxdot^1$}
\put(35.3,287){$\odot^1$}
\put(35.3,385.7){$\boxdot^1$}

\put(75.3,7){$\odot^1$}
\put(75.3,105.7){$\boxdot^1$}
\put(75.3,207){$\odot^1$}
\put(75.3,305.7){$\boxdot^1$}
\put(75.3,407){$\odot^1$}
\put(75.3,87){$\odot^1$}
\put(75.3,185.7){$\boxdot^1$}
\put(75.3,287){$\odot^1$}
\put(75.3,385.7){$\boxdot^1$}

\put(115.3,7){$\odot^1$}
\put(115.3,105.7){$\boxdot^1$}
\put(115.3,207){$\odot^1$}
\put(115.3,305.7){$\boxdot^1$}
\put(115.3,407){$\odot^1$}
\put(115.3,87){$\odot^1$}
\put(115.3,185.7){$\boxdot^1$}
\put(115.3,287){$\odot^1$}
\put(115.3,385.7){$\boxdot^1$}

\put(155.3,7){$\otimes^3$}
\put(155.3,105.7){$\boxtimes^3$}
\put(155.3,207){$\otimes^3$}
\put(155.3,305.7){$\boxtimes^3$}
\put(155.3,407){$\otimes^3$}
\put(155.3,87){$\otimes^3$}
\put(155.3,185.7){$\boxtimes^3$}
\put(155.3,287){$\otimes^3$}
\put(155.3,385.7){$\boxtimes^3$}

\put(35.3,31.22){$\times^2$}
\put(35.3,131.22){$\times^2$}
\put(35.3,231.22){$\times^2$}
\put(35.3,331.22){$\times^2$}
\put(35.3,431.22){$\times^2$}
\put(35.3,62.78){$\times^2$}
\put(35.3,162.78){$\times^2$}
\put(35.3,262.78){$\times^2$}
\put(35.3,362.78){$\times^2$}

\put(75.3,28.5){$\times^2$}
\put(75.3,128.5){$\times^2$}
\put(75.3,228.5){$\times^2$}
\put(75.3,328.5){$\times^2$}
\put(75.3,428.5){$\times^2$}
\put(75.3,65.5){$\times^2$}
\put(75.3,165.5){$\times^2$}
\put(75.3,265.5){$\times^2$}
\put(75.3,365.5){$\times^2$}

\put(115.3,21.12){$\times^2$}
\put(115.3,121.12){$\times^2$}
\put(115.3,221.12){$\times^2$}
\put(115.3,321.12){$\times^2$}
\put(115.3,421.12){$\times^2$}
\put(115.3,72.88){$\times^2$}
\put(115.3,172.88){$\times^2$}
\put(115.3,272.88){$\times^2$}
\put(115.3,372.88){$\times^2$}


\put(244,457.5){$\frac{E}{6c}$}
\put(436,-2.5){$m$}

\put(250,0){\vector(0,1){450}}
\put(250,0){\vector(1,0){180}}

\put(247,100){\line(1,0){6}}
\put(247,200){\line(1,0){6}}
\put(247,300){\line(1,0){6}}
\put(247,400){\line(1,0){6}}

\put(290,-3){\line(0,1){6}}
\put(330,-3){\line(0,1){6}}
\put(370,-3){\line(0,1){6}}
\put(410,-3){\line(0,1){6}}

\put(287.5,-12.5){$0$}
\put(327.5,-12.5){$1$}
\put(367.5,-12.5){$2$}
\put(407.5,-12.5){$3$}

\put(240,-2,5){$0$}
\put(240,97.5){$1$}
\put(240,197.5){$2$}
\put(240,297.5){$3$}
\put(240,397.5){$4$}

\put(285,10){\dashbox{2}(15,21)}
\put(325,76.6){\dashbox{2}(15,21)}
\put(365,143.3){\dashbox{2}(15,21)}
\put(405,210){\dashbox{2}(15,21)}

\put(285.3,20.33){$\odot^1$}
\put(285.3,40.33){$\odot^1$}
\put(285.3,139.03){$\boxdot^1$}
\put(285.3,240.33){$\odot^1$}
\put(285.3,339.03){$\boxdot^1$}
\put(285.3,440.33){$\odot^1$}
\put(285.3,120.33){$\odot^1$}
\put(285.3,219.03){$\boxdot^1$}
\put(285.3,320.33){$\odot^1$}
\put(285.3,419.03){$\boxdot^1$}

\put(285.3,11.79){$\times^2$}
\put(285.3,48.87){$\times^2$}
\put(285.3,148.87){$\times^2$}
\put(285.3,248.87){$\times^2$}
\put(285.3,348.87){$\times^2$}
\put(285.3,448.87){$\times^2$}
\put(285.3,111.79){$\times^2$}
\put(285.3,211.79){$\times^2$}
\put(285.3,311.79){$\times^2$}
\put(285.3,411.79){$\times^2$}

\put(325.3,86.99){$\odot^1$}
\put(325.3,106.99){$\odot^1$}
\put(325.3,205.69){$\boxdot^1$}
\put(325.3,306.99){$\odot^1$}
\put(325.3,405.69){$\boxdot^1$}
\put(325.3,186.99){$\odot^1$}
\put(325.3,285.69){$\boxdot^1$}
\put(325.3,386.99){$\odot^1$}

\put(325.3,78.45){$\times^2$}
\put(325.3,115.53){$\times^2$}
\put(325.3,215.53){$\times^2$}
\put(325.3,315.53){$\times^2$}
\put(325.3,415.53){$\times^2$}
\put(325.3,178.45){$\times^2$}
\put(325.3,278.45){$\times^2$}
\put(325.3,378.45){$\times^2$}

\put(365.3,153.66){$\odot^1$}
\put(365.3,173.66){$\odot^1$}
\put(365.3,272.36){$\boxdot^1$}
\put(365.3,373.66){$\odot^1$}
\put(365.3,253.66){$\odot^1$}
\put(365.3,352.36){$\boxdot^1$}

\put(365.3,145.12){$\times^2$}
\put(365.3,182.20){$\times^2$}
\put(365.3,282.20){$\times^2$}
\put(365.3,382.20){$\times^2$}
\put(365.3,245.12){$\times^2$}
\put(365.3,345.12){$\times^2$}
\put(365.3,445.12){$\times^2$}

\put(405.3,220.33){$\odot^1$}
\put(405.3,240.33){$\odot^1$}
\put(405.3,339.03){$\boxdot^1$}
\put(405.3,440.33){$\odot^1$}
\put(405.3,320.33){$\odot^1$}
\put(405.3,419.03){$\boxdot^1$}

\put(405.3,211.79){$\times^2$}
\put(405.3,248.87){$\times^2$}
\put(405.3,348.87){$\times^2$}
\put(405.3,448.87){$\times^2$}
\put(405.3,311.79){$\times^2$}
\put(405.3,411.79){$\times^2$}

}
\end{picture}
  \end{center}
\caption{%
The spectra of $\mu$  for various $\xi$ (left), and the total energy $E$ for $m = 0, 1, 2, 3$ (right).
The spectra of $\mu$ are plotted for $\xi = 0, {1\over 3}, {2\over 3}, 1$ under $\nu=0.9$, which interpolate between $U = V^{-1}\sigma_3 V$ and $U = \sigma_3$.
The energy spectrum of $E$ (in units of $6c$) is obtained for $\nu=0.9$ and $\xi=\frac{5}{6}$.
The signs inside the dashed squares  for the lowest two levels to each $m$ in the energy spectrum of $E$ represent the levels 
arising for
$\kappa(\lambda)=\infty$, which is allowed only for these two levels.  All other levels are obtained for the choice $\kappa(\lambda)= 0$.   Symbols used to plot the levels show the representations of the mirror-$S_3$ group;  the symbol $\odot$ stands for the identity representation $\chi^+$,  
$\boxdot$ for the signature representation $\chi^-$, and
$\times$ for the 2-dimmensioal representation $\chi^{(2)}$, respectively.  (The superimposed symbols
$\otimes$ and $\boxtimes$ indicate that the levels are degenerate by the corresponding distinct representations.)
The numbers $1, 2, 3$ on the upper right of the symbols show the total multiplicities of the levels.
}
\label{spectrum}
\end{figure}
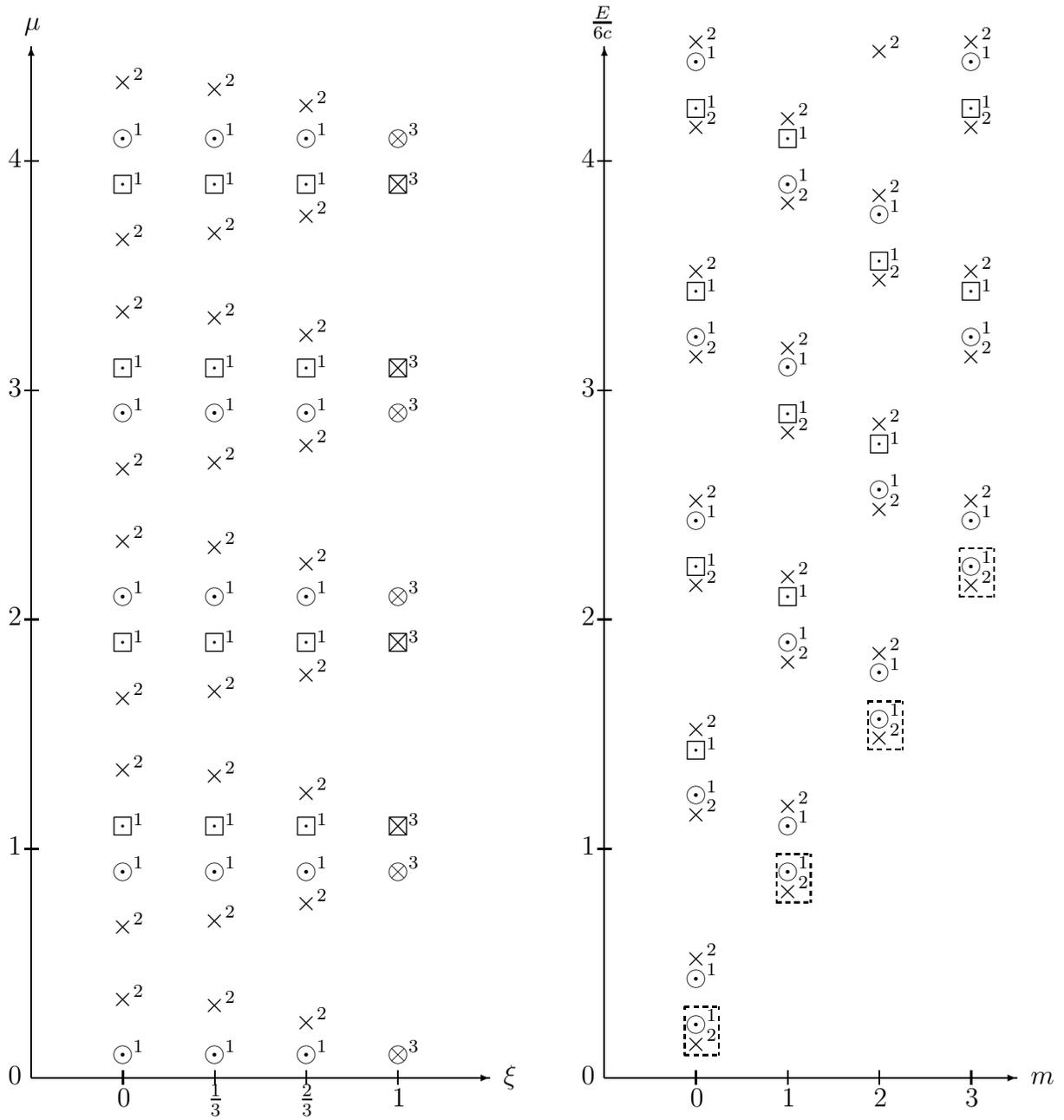

For illustration, we show in Fig.\ref{spectrum} the angular spectrum (left) for various values of 
$\xi$, and the total energy spectrum (right) for different choices of the $\kappa(\lambda)$ parameters.
We observe that if $\kappa(\lambda) = 0$ for all $\lambda$ the energy levels are made of one single series, consisting of a regular pattern formed by four singlets and four doublets.   The equispaced levels suggest that, for this case, the 
Calogero model may be solved by use of the ladder operator, 
as demonstrated for the special case \cite{Polychronakos1, Brink-Hansson-Konstein-Vasiliev} which amounts to $U=-I$ and $\kappa(\lambda)=0$ in our scheme.

\section{Spectral preserving $SU(2)$ and the angular spectrum}

The angular spectrum obtained in section 3 is independent of 
the phase of the parameter $\zeta$ even though the eigenfunctions are dependent on it.   
In particular, the levels 
of the mirror-$S_3$ singlets depend on neither of the parameters $\xi$ and $\zeta$ and hence are independent on the choice of $V$ in $U = V \sigma_3 V^{-1}$.  In fact, 
these are a consequence of the spectral preserving $SU(2)$ (or its subgroup $U(1)$) transformations \cite{Tsutsui4} which are found in the family of inequivalent quantizations we are considering here.  

To see this, let us first consider the reflection transformation $Q_1$ given by $Q_1 := P_3$ in (\ref{defprity}).   Observe that the action 
(\ref{defgection}) on the states,
\begin{align}
\psi(\phi) \to (Q_1\psi)(\phi) =  \psi(- \phi),
\end{align}
is {\it spectrum-preserving} in the sense that if $\psi_\lambda$ is an eigenstate of the angular operator 
$H_\Omega$ as in (\ref{angular equation}) with
eigenvalue $\lambda$, then 
\begin{align}
H_\Omega\,(Q_1 \psi_\lambda) (\phi) = \lambda \, (Q_1 \psi_\lambda) (\phi), 
\label{spone}
\end{align}
on account of the formal invariance of the operator $H_\Omega$ under $Q_1$.  Note that 
this does not imply that 
$Q_1$ is a symmetry unless it is compatible with the boundary conditions specified by $U$.  For the boundary vectors, we find
\begin{align}
B_{k}(Q_1\psi)   =   \sigma_1 B_{6-k}(\psi) , \qquad
B_{k}'(Q_1\psi)   =    \sigma_1 B_{6-k}'(\psi),
\label{bcone}
\end{align}
for $k = 0$, $\ldots$, 5, with the identification $B_{0}=B_{6}$ and $B'_{0}=B'_{6}$.  Thus, in effect, the 
transformation $Q_1$ induces in the connection conditions (\ref{cconk}) the change
\begin{align}
U \to \sigma_1\, U \, \sigma_1.
\end{align}
It follows that the operator $H_\Omega$ shares the same spectrum under the boundary conditions specified by $U$ and 
$\sigma_1\, U \, \sigma_1$.

We next consider
the \lq alternate reflection\rq\ defined by
\begin{align}
\psi (x)
\to (Q_3 \psi)(\phi) :=
\left\{
\begin{array}{ll}
\psi(\phi), & \phi_{2k} < \phi < \phi_{2k+1}, \\
-\psi(\phi), & \phi_{2k+1} < \phi < \phi_{2k+2}. 
\end{array}     
\right.               
\label{spthree}
\end{align}
Evidently, this is also spectral-preserving, and on the boundary vectors we have
\begin{align}
B_{k}(Q_3\psi)   =   \sigma_3 B_{k}(\psi) , \qquad
B_{k}'(Q_3\psi)   =    \sigma_3 B_{k}'(\psi).
\label{bcthree}
\end{align}
Accordingly,  we find that the alternate reflection (\ref{spthree}) induces 
\begin{align}
U \to \sigma_3\, U \, \sigma_3
\end{align}
in the connection conditions (\ref{cconk}).  From  $Q_1$ and $Q_3$ we further define the product transformation $Q_2 := iQ_1 Q_3$ that yields
\begin{align}
B_{k}(Q_2\psi)   =   \sigma_2 B_{6-k}(\psi) , \qquad
B_{k}'(Q_2\psi)   =    \sigma_2 B_{6-k}'(\psi).
\label{bctwo}
\end{align}
By construction, $Q_2$ is spectral-preserving and induces
\begin{align}
U \to \sigma_2\, U \, \sigma_2.
\end{align}
Consequently, we see that the connection conditions by
$U$ and $\sigma_i\, U \, \sigma_i$ for all $i = 1, 2, 3$ give rise to the same spectrum
for the operator $H_\Omega$.  Note that 
$Q_i$'s form the $su(2)$ algebra,
\begin{align}
\left[Q_i, \, Q_j \right]  = 2i\epsilon^{ijk}Q_k.
\label{sutwo}
\end{align}

Suppose, now, that the state $\psi$ is in a singlet representation of the mirror-$S_3$, {\it i.e.}, it is an eigenstate of $R_i$ with fixed eigenvalues $\pm 1$ for all $i = 1, 2, 3$. 
The state $\psi$ is then a periodic function with period ${2\pi}/3$, and hence we have 
$B_{k}(\psi) =B_{-k}(\psi)$ and $B'_{k}(\psi) =B'_{-k}(\psi)$.  
The three relations (\ref{bcone}), (\ref{bcthree}) and (\ref{bctwo}) can then be combined as
\begin{align}
B_{k}(Q_i\psi)   =   \sigma_i B_{k}(\psi) , \qquad
B_{k}'(Q_i\psi)   =    \sigma_i B_{k}'(\psi),  \qquad i = 1, 2, 3.
\label{sptrb}
\end{align}
On account of the linearity of $B_{k}$ and $B'_{k}$ observed in (\ref{sptrb}) and the algebraic property (\ref{sutwo}),  we find that 
a linear combination of the $Q_i$'s
\begin{align}
Q = \sum_{i=1}^3 c_i Q_i, \qquad  \sum_{i=1}^3 c_i^2 = 1,
\label{gentrb}
\end{align}
with arbitrary  coefficients $c_i$ fulfills $Q^2 = I$ and 
is also spectral-preserving for singlet states.   On the matrix $U$, this induces 
\begin{align}
U \to \sigma\, U \, \sigma, 
\qquad
\sigma = \sum_{i=1}^3 c_i \,\sigma_i.
\label{genindmat}
\end{align}
It can be shown \cite{Tsutsui4} that by choosing $c_i$ appropriately one finds $\sigma$ such that
$\sigma\, U \, \sigma = V^{-1} U V$ for any $V \in SU(2)$.  In other words, the transformation generated by
$Q$ in (\ref{gentrb}) yields the change
\begin{align}
U \to V^{-1}\, U \, V,
\label{transformation of u}
\end{align}
without altering the spectrum.   It then follows that, as far as the mirror-$S_3$ singlets are concerned, the operator $H_\Omega$ shares the same spectrum under
$V^{-1} U V$ for any $V \in SU(2)$.  From this we see that for the scale invariant $U = V \sigma_3 V^{-1}$ 
we are considering, the spectrum of the mirror-$S_3$ singlets is independent of $V$, as we have observed in the last section.  

This $SU(2)$-independence does not hold for the eigenstates in the doublet representation $\chi^{(2)}$ of the mirror-$S_3$, because for them the transformations $Q_1$ and $Q_2$ interchange different boundary vectors as seen in (\ref{bcone}) and (\ref{bctwo}).   However, the transformation $Q_3$ still maps the boundary vectors to themselves as (\ref{bcthree}) and can be used to provide a 1-parameter family of spectral-preserving transformations
$e^{i\theta Q_3} $ for any real $\theta$.  On the matrix $U$, the transformations induce
\begin{align}
U \to e^{-i\theta \sigma_3}\, U \, e^{i\theta \sigma_3},
\end{align}
which form a $U(1)$ subgroup of the $SU(2)$ transformations (\ref{transformation of u}).
In parameters, this allows us to alter the phase of $\zeta$ in $U$ freely, which implies that 
the whole spectrum depends only on $|\zeta|$ or $\xi$ as seen earlier.

Finally, we mention that the universality in the spectrum is a general feature of a 
circle system with even number $2N$ of singularities that appears when it is quantized under mirror symmetries defined analogously to the present case $2N = 6$.

\section{Conclusion}

In the present paper we studied the inequivalent quantizations of the $N = 3$ Calogero model based on the method of separation of variables.  Our inequivalent quantizations respect both the mirror-$S_3$ invariance and the scale invariance.   These quantizations  are, in a sense, supplemental to the quantizations presented in the paper 
\cite{Tsutsui2} which respect the $D_6$ invariance, in view of the fact that  
the $D_6$ is restored when the scale invariance is exchanged for parity invariance in our case.  Our symmetry requirement is that all the connection conditions at the singularities in the angular part are specified by a single matrix $U$ of the form
(\ref{radial boundary condition2}).  For a system consisting of a line, this class of singularities
is known as the scale invariant family and is given, except for the cases $U = \pm {\bf 1}_2$, by an $S^2$ parameter space (see (\ref{umat})).  We mention that the scale invariant family supports the Berry phase (or anholonomy) when the singularities are tuned along a cycle on the scale invariant sphere $S^2$ \cite{Tsutsui5}.

In our inequivalent quantizations, we found the eigenstates and eigenvalues explicitly, both for the angular and radial parts that arise after the
separation of variables is made.  These eigenstates are classified in terms of the irreducible representations of the mirror-$S_3$ group.  We observed that the eigenvalues corresponding to the singlets of the $S_3$ are independent of the choice of $U$ in the family, whereas those corresponding to the doublets of the $S_3$ are dependent only on one parameter $\xi$ which corresponds to the coordinate along a great circle on the $S^2$.   We showed that these properties are due to the spectral-preserving $SU(2)$ transformations or $U(1)$ transformations that the scale invariant family possesses.

The scale invariance is strict enough to narrow the possible boundary conditions  at $r = 0$ 
in the radial part down to just
two types, one given by the Dirichlet condition and the other by the Neumann condition.   The Neumann condition is possible if $\lambda < 1$, which occurs only for the eigenstates coupled to the lowest two angular levels.    The possibility of the Neumann condition brings about a number of different spectra for the total energy, as we have seen in Fig.\ref{spectrum}.   When all the radial eigenstates adopt the Dirichlet condition irrespective of the eigenvalue $\lambda$ of the radial level that couples to them,  the energy spectrum exhibits a regular pattern consisting of a number of distinct sets of levels which are equispaced from each other.   This suggests that for these cases we may also devise the method of the ladder operator to solve the model.  In contrast, this seems to be impossible when the Neumann condition is adopted for the lower levels, where the regular pattern is broken at the lowest end of the spectrum.   The appearance of the intriguing combination of inequivalent quantizations, which we found in the angular and radial parts in our model, is perhaps a generic feature to be observed in quantizing models of more than two dimensions in general frameworks.

\bigskip
\bigskip
\noindent{\bf Acknowledgements.}
One of the authors (I.T.) thanks Professor L. Feh\'er and Dr. T. F\"ul\"op for helpful discussions.
This work was supported by the Grant-in-Aid for Scientific Research, 
No.~13135206 and No.~16540354, of the Japanese Ministry of Education, 
Science, Sports and Culture.

\newpage



\end{document}